\documentstyle[emulateapj,pstricks]{article}
\def\msun{\mbox{M$_\odot$}}
\def\I{{\'\i}}

\begin{document}
\slugcomment{{\em submitted to the Astrophysical Journal}}
\lefthead{CHEMICAL EVOLUTION OF IRREGULAR GALAXIES}
\righthead{CARIGI ET AL.}

\title{DARK MATTER AND THE CHEMICAL EVOLUTION OF IRREGULAR GALAXIES}

\author{L. Carigi, P. Col\I n, and M. Peimbert}
\affil{Instituto de Astronom\I a, Universidad Nacional Aut\'onoma
de M\'exico, Mexico}

\begin{abstract}

We present three types of chemical evolution models for irregular
galaxies: closed-box with continuous star formation rates (SFRs), 
closed-box with bursting SFRs, and O-rich outflow  with continuous SFRs.
We discuss the chemical evolution of the irregular galaxies NGC 1560 and 
II Zw 33, and a ``typical'' irregular galaxy. The fraction of low-mass 
stars needed by our models is larger than that derived for the solar 
vicinity, but similar to that found in globular clusters. For our typical 
irregular galaxy we need a mass fraction of about 40 \% in the form of 
substellar objects plus non baryonic dark matter inside the Holmberg radius,
in good agreement with the results derived for NGC 1560 and II Zw 33 where 
we do have an independent estimate of the mass fraction in non baryonic dark 
matter. Closed-box models are better than O-rich outflow models in explaining 
the C/O and $Z/{\rm O}$ observed values for our typical irregular galaxy. 
\end{abstract}
\keywords{galaxies: abundances -- galaxies: evolution --
galaxies: irregular -- stars: luminosity function, mass function}

\section{INTRODUCTION}

\begin{planotable}{lccccc}
\tablewidth{0pt}
\tablecaption{$r$ Values }
\tablehead{\colhead {Object} & \colhead {Ref} &
\colhead{$\alpha_1$\tablenotemark{a}} & \colhead{0.01-85 \msun} &
\colhead{0.01-120 \msun} & \colhead{0.08-85 \msun}} 
\startdata
Solar vicinity & KTG & 1.30 & 1.000 & 1.000 & 1.000 \nl
Solar vicinity & KTG & 1.85 & 1.512 & 1.510 & 1.199 \nl
Solar vicinity & KTG & 0.70 & 0.820 & 0.820 & 0.883 \nl
NGC 6752       &FCBRO& 1.90, 2.33 \tablenotemark{b}
                            & 1.879 & 1.876 & 1.368 \nl
NGC 7099       & PCK & 2.00 & 1.812 & 1.809 & 1.279 \nl
NGC 6397       & PCK & 1.60 & 1.202 & 1.201 & 1.093 \nl
\enddata
\tablenotetext{a}{
$\rm{IMF} \propto \cases{m^{-\alpha_1} & if $m_l \leq m < 0.5$ , \cr
                         m^{-2.2}      & if $0.5 \leq m < 1.0$ , \cr
                         m^{-2.7}      & if $1.0 \leq m < m_u$. \cr}$ 
  }
\tablenotetext{b}{
$m^{-1.9}$ if $m_l \leq m < 0.27$, 
\hskip 0.5cm
$m^{-2.33}$ if $0.27  \leq m < 0.5$. 
  }
\end{planotable}

Carigi et al. (1995, hereinafter CCPS) presented a series of models of
chemical evolution of irregular galaxies; they concluded that an IMF
with a larger fraction of low-mass stars than in the solar vicinity,
as well as the presence of a moderate O-rich outflow, were needed to fit a
series of observational constraints provided by a ``typical" irregular
galaxy, these constraints correspond to average values of well observed
irregular galaxies. The models were computed based
on  the yields by Maeder (1992) and under the assumption that no non baryonic
matter was present. Recent advances led us to
produce a new set of models taking into account the following
developments: a) the determination of accurate IMFs for globular
clusters based on HST observations, b) the determinations of new yields
for massive stars by Woosley, Langer \& Weaver (1993) (WLW) and
Woosley \&  Weaver (1995) (WW), c) the determination of the
amount of dark matter in several irregular and spiral galaxies, 
d) better estimates of the H$_2$ content in irregular galaxies.

In \S 2 we define the different mass fractions present in an irregular
galaxy and their role in chemical evolution models.
In \S 3 we discuss the evidence in favor of a larger fraction of low-mass 
stars in the IMF than previously adopted for the solar vicinity,
consequently we present modified IMFs characterized by  a parameter $r$ also
defined in this section.
In \S 4 we present the assumptions adopted for our chemical evolution models.
In \S 5 we present models for two irregular galaxies, NGC 1560 and II Zw 33, 
for which the amount of non baryonic dark matter is known.
In \S 6 we present the observational properties of a typical irregular
galaxy  and produce closed-box models with continuous SFRs
that reproduce the observational
constraints.
We have also computed closed-box models with bursts of star formation
to compare them with the continuous SFR models.
The discussion and conclusions are presented in \S 7 and \S 8, respectively.

\section {DEFINITIONS}

We can define the gas mass fraction of a galaxy, $\mu$, as

\begin{equation}
 \mu = M_{gas}/M_{total}=M_{gas}/(M_b+M_{nb})
\end{equation}

\noindent
where $M_b$ is the baryonic mass and $M_{nb}$ is the non baryonic mass
(we are defining as ``$M_{nb}$" the matter that does not follow the
stellar or gaseous mass distribution),
often $M_{nb}$ has been called the dark halo mass. $M_b$ can be expressed as

\begin{equation}
M_b = M_{gas} + M_{sub} + M_{vl} + M_{rest} + M_{rem}
\end{equation}

\noindent
where $M_{sub}$ is the mass in substellar objects ($m<0.1$, throughout
this paper $m$ is given in solar masses)
(we are also defining ``$M_{sub}$" as the baryonic dark matter),
$M_{vl}$ is the mass of stars in the $0.1 \leq m < 0.5$ range,
$M_{rest}$ is the mass of stars in the $0.5 \leq m \leq 85$ range, and
$M_{rem}$ is the mass in compact stellar remnants.
It is possible to evaluate $\mu$ observationally {from} a direct determination
of $M_{gas}$ and a dynamical determination of $M_{total}$.

To produce a chemical evolution model we need to reach a certain O abundance
at a given gas mass fraction of the material that participates
in the chemical evolution process, $\mu_{IMF}$, given by

\begin{equation}
\mu_{IMF}= M_{gas}/M_b
\end{equation}

\noindent
which together with equation (1) yields

\begin{equation}
\mu_{IMF}=\mu (1 + M_{nb}/M_b)
\end{equation}

{From} an observational value of $\mu$, but without knowing $M_{nb}/M_b$,
{from} equation (4) it follows that we can only derive a lower limit for
$\mu_{IMF}$.
Note that in O-rich outflow models $\mu_{IMF}$ does not include the 
mass lost in the outflow.

\section{THE $r$ PARAMETER}

\subsection{An Initial Mass Function}

The IMF for the solar neighborhood adopted in this paper is
what was called KTG IMF (Kroupa et al. 1993) in CCPS, except that here the upper limit
is taken to be 85 \msun, and it is given by
\begin{equation}
\xi (m) = \cases{ 0.506~m^{-1.3} & if $0.01 \leq m < 0.5$ , \cr
                    0.271~m^{-2.2} & if $0.5 \ \leq m < 1.0$ , \cr
                    0.271~m^{-2.7} & if $1.0 \ \leq m \leq 85$ , \cr}
\end{equation}
where $ \xi(m)dm$ is the number of stars in the mass interval {from} 
$m$ to $m + dm$. This function is extended to 
a minimum mass of 0.01  to take into account the fraction of
dark matter hidden in objects of substellar mass.
This function is normalized to one; that is,
\begin{equation}
\int_{m_l}^{m_u} m \xi(m) dm = 1,
\end{equation}
where $m_l$ and $m_u$, the lower and upper limits, are taken to be 0.01  and 85,
respectively, unless otherwise stated.

\subsection{Chemical Evolution Models and the Value of $r$ }

An $r$ value was defined in CCPS as follows
\begin{equation}
r(m_l,m_u) = {\int_{0.5}^{m_u}m\xi(m)dm \over \int_{0.5}^{m_u} m\xi'(m)dm},
\end{equation}
where $m_l$ and $m_u$ are the low-mass and the high-mass end of the
IMF respectively, $\xi(m)$ is the KTG IMF, and 
$\xi'(m)$ is an IMF with different slope in $m_l \leq m < 0.5$ range 
(see equation 5 and Table 1); 
$r$ depends on $m_l$ because 
$\xi(m)$ and $\xi'(m)$ are normalized to one (see equation 6).

The chemical evolution models depend on $r$ because the net yields of the
heavy elements, and in particular that of O, decrease when $r$ increases.
(See the definition of net yield in Peimbert, Col\I n \& Sarmiento 1994,
and models for different $r$ values in CCPS).

The closed-box model
based on the yields by Maeder (1992) is unable to
reproduce simultaneously the $\mu$, O, C/O, $Z$/O, and 
$\Delta Y/\Delta {\rm O}$ values of the typical irregular galaxy studied by CCPS.
CCPS were able to reproduce the observed constraints based on an O-rich
outflow model with $\gamma=0.23$ and $r(0.01,120)=2.66$, where
$\gamma$ is the fraction of O produced by SNe that is ejected to the 
intergalactic medium without mixing with the interstellar gas.

We want to study the effect of using different yields in the models.
We will produce closed-box models for different $r$
values to fit the observational constraints.
We will discuss if these $r$ values are in agreement with the observed IMFs 
of globular clusters and of the solar vicinity.

\subsection{Solar Vicinity}

We will study two problems: the $r$ values determined for different IMFs and
the effect of the yields by Maeder and WLW \& WW on the $M_{sub}$ value
for chemical evolution models of the solar vicinity.

In Table 1 we present $r$ values for different mass ranges and for different
values of the slope for the low-mass range, $\alpha_1$, given by 
${\rm IMF} \propto m^{-\alpha_1}$
for $m_l \leq m < 0.5$; we have adopted the IMF slopes given by KTG for $m \geq 0.5$.
In Table 2 we present the average mass of the objects in the IMF
{from} $m_x$ to $m_u$, when $m_x=m_l=0.01$ we include all objects in the
IMF and when $m_x=0.1$ we include only stellar objects in the IMF.

The KTG simulations of star-count data reach the maximum confidence when
the scale height for their model reaches 270 pc; in this case $\alpha_1 = 1.3$.
Based on their model KTG suggest for the solar vicinity that $0.70 < \alpha_1< 1.85$,
for the $0.08 \leq m < 0.5$ range. We adopted $\alpha_1 = 1.3$ as the preferred value, and
for this value we define $r = 1$ for the three mass ranges presented in Table 1 (see
the first line of this table).

We present three mass ranges in Table 1 for the following reasons: Maeder (1992),
and WLW (1993) use $m_u$ equal to 120 and 85 respectively,
while CCPS and in this paper we have adopted $m_l = 0.01$, in the last column we
present the lower mass included by KTG, $m_l = 0.08$.

Carigi (1996) computed a chemical evolution model of the solar
neighborhood with $M_{nb}=0$ based on the yields by Maeder (1992) adopting the KTG
IMF for the $0.01 \leq m \leq  120$ range. For this model the mass fraction in
substellar objects amounts to 15.1 \% and the present
day gas fraction is 0.15 in agreement with the data (0.05-0.20, Tosi 1996).

We have computed a model for the solar neighborhood under the same assumptions
as those adopted by Carigi (1996), but based on the yields by WLW and WW,
and adopted a $0.01 \leq m \leq 85$ mass range ($r=1$). We
obtain for this model a mass fraction in substellar objects of 14 \% and 
the present day gas fraction is 0.07, also in agreement with the observations
(Tosi 1996).

Both models of the solar vicinity could accommodate an additional
modest amount of substellar objects or non baryonic
dark matter and still reproduce the observed abundances and the
$\sigma_{gas}/\sigma_{total}$ value.

If there is a significant fraction of non baryonic dark matter that does
not participate in the chemical evolution process, it has to be subtracted
{from} $M_{total}$ increasing the value of $\mu_{IMF}$. 
Therefore to reach the observed
abundances with lower gas consumption the IMF needs to have a larger fraction
of massive stars, and if we keep the KTG slope for massive stars constant we
need to reduce the fraction of low-mass stars in the model, and consequently
$r$ values smaller than one.

On the other hand, if $M_{total}$ is larger than the observed
value adopted by us, and if this difference is due to a greater mass
fraction in substellar objects in the solar vicinity, then the $\mu_{IMF}$ would become
smaller and the model would need a higher gas consumption and $r$ values higher
than one to be able to reproduce the observed abundances.
\begin{planotable}{cccccc}
\tablecolumns{6}
\tablewidth{0pt}
\tablecaption{$<m>$ Values }
\tablehead{\colhead{$\alpha_1$\tablenotemark{a}} &
\colhead{0.1-85 \msun} & \colhead{0.1-120 \msun} &
\colhead{0.01-85 \msun} &
\colhead{0.01-120 \msun} & \colhead{0.08-85 \msun}}
\startdata
 1.30     & 0.501 & 0.503 & 0.194 & 0.195 & 0.452 \nl
 1.85     & 0.391 & 0.393 & 0.074 & 0.074 & 0.335 \nl
 0.70     & 0.626 & 0.629 & 0.429 & 0.431 & 0.593 \nl
1.90, 2.33& 0.345 & 0.346 & 0.062 & 0.062 & 0.293 \nl
 2.00     & 0.364 & 0.366 & 0.058 & 0.058 & 0.307 \nl
 1.60     & 0.440 & 0.441 & 0.116 & 0.116 & 0.386 \nl
\enddata
\tablenotetext{a}{ As in Table 1.}
\end{planotable}

\subsection{Globular Clusters}

Based on HST observations Ferrano et al. (1997, hereinafter FCBRO) determined
the mass function for the lower main sequence of the globular cluster NGC 6752
($Z = 0.03 \ Z_\odot$). They found $x$ values of 0.90 and 1.33 for the mass ranges
$0.15 - 0.30$ \msun \ and $0.25 - 0.55$ \msun, respectively, 
where $\xi'$ is proportional to
$m^{-(1 + x)}$. Therefore we have adopted for NGC 6752:

\begin{equation}
\xi'_{6752} (m) \propto \cases{ m^{-1.9 } & if $m_l   \leq m < 0.27$ , \cr
                        m^{-2.33} & if $0.27  \leq m < 0.5 $ , \cr
                        m^{-2.2 } & if $0.5 \ \leq m < 1.0 $ , \cr
                        m^{-2.7 } & if $1.0 \ \leq m \leq m_u$. \cr}
\end{equation}

Note that to derive an $r$ value we need to define the IMF in the
$m_l$-$m_u$ range and that for $m \geq 0.5$ we are adopting KTG for all
globular clusters.

Also based on HST observations Piotto, Cool, \& King (1997, hereinafter PCK) have
determined the mass function of four globular clusters. Three of them have very
similar mass functions: NGC 6341, NGC 7078 and NGC 7099. Based on the mass
luminosity relation by D'Antona and Mazzitelli (1995) PCK derive for NGC 7099
that $x = 1.0$ for masses below 0.4 \msun. Consequently, we have adopted for
NGC 7099:

\begin{equation}
\xi'_{7099} (m) \propto \cases{ m^{-2.0} & if $m_l  \leq m < 0.5 $ , \cr
                        m^{-2.2} & if $0.5 \ \leq m < 1.0 $ , \cr
                        m^{-2.7} & if $1.0 \ \leq m \leq m_u$. \cr}
\end{equation}

For the same mass range, and the same mass luminosity relation PCK find $x=0.6$
for NGC 6397. They suggest that the lower value of $x$ derived for NGC 6397,
relative to the other three clusters, could be due to selective loss of low-mass
stars by evaporation and tidal shocks.

In Table 1 we present the NGC 6752, NGC 7099 and NGC 6397 $r$ values for the
different mass ranges considered. By comparing the $r$ values for different mass
ranges it is found that the effect on $r$ introduced by changing the upper mass
end {from} 120 to 85 \msun \ is negligible. Moreover,
 it is also found that the $r$ values
for NGC 6752 and NGC 7099 are significantly higher than for the solar vicinity.
This result implies that the fraction of low-mass stars is higher for globular
clusters than for the solar vicinity, and might imply that the fraction
of substellar objects is higher also.

\section{ CHEMICAL EVOLUTION MODELS}

All the models in this paper reproduce at least two observational 
constraints:
the O abundance by mass in the interstellar medium (ISM), and $\mu$.
In addition each model predicts different element abundance ratios
by mass
that can be compared with other observational constraints.

We computed three types of models:
closed-box with continuous SFRs, closed-box with bursting SFRs,
and O-rich outflow  with continuous SFRs.
The assumptions adopted
in our models are:

a) The initial composition of the gas is primordial: $Y_0=0.23$, $Z_0=0.00$.

b) We have computed models for three galaxy ages, $t_g$: 0.1, 1.0, and 10.0 Gyr

c) The star formation rate is proportional to the gaseous mass, 
SFR = $\nu M_{gas}$. The efficiency,
$\nu$,
is mainly determined by the need to reach $\mu_{IMF}$
at the age of the model.
For a continuous SFR $\nu$ is constant in time.
On the other hand,  when we consider a bursting SFR
\begin{equation}
\nu = \cases{ constant & if $t_j               \leq t < t_j + 40$  Myr ,\cr
              0        & if $t_j + 40 \ {\rm Myr} \leq t \leq t_{j+1}$, \cr}
\end{equation}
where $t_j = (t_g - 0.2) {(j-1) \over (n-1)}$ Gyr is the burst starting time,
$n$ is the total number of bursts, and $1 \leq j \leq n$.

d) We have adopted several IMFs (see \S 3). For $m \geq 0.5$ all of them have
 the
same slopes as those given by KTG. The $r(0.01,85)$
 value is varied until
the desired oxygen abundance is obtained. This $r(0.01,85)$ value corresponds to a unique
IMF
with a slope for stars with $m < 0.5$ denoted by $\alpha_1$.
In what follows $r$ corresponds to the 0.01-85 mass interval
unless otherwise noted.

e) We drop the instantaneous recycling approximation, IRA, and assume that the 
stars
eject their envelopes after leaving the main sequence. The main sequence 
lifetimes
are taken {from} Schaller et al. (1992). The possible reduction of the O yields of
massive stars due to the production of black holes as suggested by Maeder (1992)
has not been considered.

f) We have used the stellar yields and remnant masses due to: 
i) Renzini \& Voli (1981) for $1.0 \leq m \leq 8.0$ ($\alpha = 1.5$, $\eta = 1/3$);
ii) WW  for $11 \leq m \leq 40$ (models ``B" for 30, 35 and 40 \msun);
iii) WLW for $ m = 60$ and $m=85$.
We also consider the changes in the stellar yields due to the stellar initial 
metallicity.
Only massive stars, those with $m > 8$, enrich the ISM with oxygen.

g) For SNIa we have taken into account the yields by Nomoto, Thielemann, \& 
Yokoi
(1984, model W7). Only a fraction of binary stars, in the $3 \leq m_1 + m_2 \leq 16$
 range,
become SNIa;
where $2.2 \leq m_1 \leq 8.0 $ and $0.8 \leq m_2 \leq 8.0 $ .
  We have determined such fraction by fitting the observed solar Fe
abundance.

h) In models with outflow of O-rich material we assume that a fraction,
$\gamma$, of the mass expelled by type II SNe is ejected to the 
intergalactic medium without 
mixing with the ISM. The WW yields have been computed
without considering stellar winds, therefore the ejected mass during 
a SN explosion is equal to the initial mass minus the stellar remnant.

\section{ NGC 1560 AND II Zw 33 }

There are two irregular galaxies with good O/H values for which 
$\mu_{IMF}$ can be determined: NGC 1560 and
II Zw 33, also known as {\sl Markarian} 1094.
For these two galaxies we can compute  closed-box models for
different ages, each model characterized by
an $r(0.01,85)$ value, which corresponds to a specific IMF.

\subsection{ $\mu_{IMF}$ and  O/H}

{From} the studies of the rotation curves of a few dwarf  irregular
galaxies it is found that they are dominated by non baryonic dark matter
(e.g., Burlak 1996, Salucci \& Persic 1997),
some of them well within the core of the mass distribution
(e.g., Moore 1994). In general  most of the $M_{nb}$ is present outside the
Holmberg radius. Unfortunately, for most of these non baryonic dominated
galaxies, for which a rotation curve is available, chemical abundance
determinations do not exist.
{From} the very reduced group for which rotation curves as well as oxygen
 abundances are available we have extracted NGC 1560 and II Zw 33 to
build chemical evolution models.
For these galaxies $M_{gas}$, $M_b$, and consequently $\mu_{IMF}$ are
known and are presented in Table 3 (see Walter et al. 1997 and
Broeils 1992).

The O/H gaseous value 
for NGC 1560 comes {from} Richer \& McCall (1995)
and for II Zw 33 comes {from} Esteban \& Peimbert (1995).
To derive the O abundances by mass we have considered the
contribution of O expected to be in dust grains (0.04 dex) and the effect
of temperature variations over the observed volume (0.16 dex); consequently
we have added 0.2 dex to the gaseous values derived under the
assumption of a constant temperature distribution inside the
H II regions (see CCPS and references therein).

\begin{planotable}{lccccc}
\tablecolumns{6}
\tablewidth{0pt}
\tablecaption{Properties of NGC 1560 and II Zw 33 \label{tab:galaxies}}
\tablehead{\colhead{Galaxy} & \colhead{$\log (M_{total}/\msun)$} &
\colhead{$\log (M_{gas}/\msun)$} & \colhead{$\log \mu$} &
\colhead{$\log \mu_{IMF}$} & \colhead{10$^3$O}}
\startdata
NGC 1560 &  9.83\phm{100} & 9.20 & $-0.63$ & $-0.34$ &  2.03 \nl
II Zw 33 &  9.71\phm{100} & 9.17 & $-0.54$ & $-0.54$ &  1.93 \nl
\enddata
\end{planotable}

\subsection { Closed-box Models with Continuous SFRs for
NGC 1560 and II Zw 33}

Closed-box  models for  NGC 1560 and II Zw 33
have been computed under the assumptions presented in \S 4.
The models reproduce $\mu_{IMF}$ and the O abundance by mass shown in Table 3.

In Table 4 we present three
models computed for NGC 1560 with  continuous SFRs.
Each line of this Table
represents a different model. For each age we find a unique
$r(0.01,85)$ value. Columns 3 to 6 show the mass fractions
defined in equations (1) and (2). The
C/O, $\Delta Y/\Delta {\rm O}$, and $Z$/O ratios are presented in
columns 7 to 9; in all tables C, O, $Y$ and $Z$ are given by mass.

The models with $\gamma=0.00$ and $M_{nb}=0.0$ for II Zw 33 
are presented in Tables 5 and 6, for these models 
$\mu_{IMF}=0.29$.
{From} these models we note that:
a) $r$ increases with model age because a larger fraction of stars have
enriched the ISM with heavy elements and the model needs to reduce $M_{rest}$
(columns 5 and 6 of Tables 4 and 5, respectively);
b) if $r$ increases, $\alpha_1$ increases and
$M_{sub}$ becomes higher;
c) $M_{vl}$ changes little with age;
d) despite the fact that $M_{rest}$ decreases with  age
$M_{rem}$ grows with age because the fraction of stars that have had
time to end their evolution is higher;
e) for NGC 1560 $M_{sub} + M_{nb} \sim 61$ \% and $r \sim 1.73$
while for II Zw 33, $M_{sub} + M_{nb} \sim 43$ \% and $r \sim 2.75$,
at 10 Gyr;
f) the 1.0-Gyr and 10-Gyr models predict C/O and $Z$/O ratios higher than
those determined observationally for our typical irregular galaxy (see
Tables 9);
g) the 0.1-Gyr models reproduce well the observed
C/O and $Z$/O ratios but predict lower $\Delta Y/\Delta {\rm O}$
values than observed.

\begin{planotable}{ccccccccc}
\tablecolumns{9}
\tablecaption{Models for NGC 1560\tablenotemark{a}}
\tablehead{ \colhead{$t_g$(Gyr)} &  \colhead{$r$} & 
\colhead{$M_{sub}(\%)$} & \colhead{$M_{vl}(\%)$} & \colhead{$M_{rest}(\%)$}
& \colhead{$M_{rem}(\%)$} & 
\colhead{C/O} & \colhead{$\Delta Y/\Delta {\rm  O}$} &\colhead{$Z$/O}}
\startdata
 0.1  & 1.471 &  8.9 &  8.7 &  9.9 & 0.2 & 0.167 & 2.844 & 1.842 \nl
 1.0  & 1.615 & 10.6 &  8.8 &  7.5 & 0.8 & 0.292 & 3.678 & 2.268 \nl
 10.0 & 1.727 & 12.0 &  9.0 &  4.5 & 2.2 & 0.324 & 4.167 & 2.429 \nl
\enddata
\tablenotetext{a}{$M_{sub} + M_{vl} + M_{rest} + M_{rem} + M_{gas} + M_{nb} = 100.0\%$, with 
$ M_{gas} = 23.4$ \%, $M_{nb}=48.9 \%$, and $\mu_{IMF}=0.46$}
\end{planotable}

\subsection { Additional Models for II Zw 33 }

There is no compelling observational evidence for large systematic IMF
variations in galaxies for $m>1.0$ (e.g., Kennicutt 1998 and references 
therein).
Therefore it is
possible that the IMF could be the same everywhere for $m < 1.0$.  
This would
imply a unique $r$ value for all objects. Therefore we
will explore if it is possible to produce chemical evolution
models for II Zw 33 with $r = 1.8$, the average value for NGC 6752,
NGC 7099, and NGC 1560 ($t_g=10$ Gyr).
There are two groups
of models that satisfy the $r = 1.8$ requirement; closed-box models with $M_{nb} \ne$ 0.0
and O-rich outflow models with $M_{nb}$ = 0.0 . In what follows we will explore these two
possibilities further.

\begin{planotable}{ccccccccc}
\tablecolumns{9}
\tablecaption{Models for II Zw 33\tablenotemark{a}}
\tablehead{ \colhead{$t_g$(Gyr)} &  \colhead{$r$}
&\colhead{$\gamma$} &
\colhead{$M_{sub}(\%)$} & \colhead{$M_{vl}(\%)$} & \colhead{$M_{rest}(\%)$}
& \colhead{$M_{rem}(\%)$} &\colhead{$M_{nb}(\%)$}
& \colhead{$\mu_{IMF}$} }
\startdata
 0.1  & 2.502 & 0.00 & 38.6 & 17.7 & 14.6 & 0.3 & 0.0  & 0.29 \nl
      & 1.800 & 0.00 & 17.6 & 12.4 & 12.5 & 0.3 & 28.4 & 0.41 \nl
      & 1.800 & 0.29 & 29.5 & 20.8 & 20.4 & 0.5 & 0.0  & 0.29 \nl
 1.0  & 2.632 & 0.00 & 41.0 & 17.7 & 11.2 & 1.3 & 0.0  & 0.29 \nl
      & 1.800 & 0.00 & 16.2 & 11.4 &  8.7 & 1.0 & 33.9 & 0.44 \nl
      & 1.800 & 0.33 & 30.7 & 21.7 & 16.8 & 2.0 & 0.0  & 0.29 \nl
 10.0 & 2.750 & 0.00 & 43.3 & 17.8 &  6.6 & 3.5 & 0.0  & 0.29 \nl
      & 1.800 & 0.00 & 15.4 & 10.9 &  5.3 & 2.6 & 37.0 & 0.46 \nl
      & 1.800 & 0.38 & 32.3 & 22.8 & 10.5 & 5.6 & 0.0  & 0.29 \nl
\enddata
\tablenotetext{a}{$M_{sub} + M_{vl} + M_{rest} + M_{rem} + M_{gas} + M_{nb} = 100.0\%$,
with $ M_{gas} = 28.8$ \%}
\end{planotable}

\subsubsection { Closed-box Models with $M_{nb} \ne$ 0 }

In Tables 5 and 6 we present the main characteristics of models with $r$ = 1.8
and $M_{nb} \ne 0.0$ for
II Zw 33. The two main differences between the models with $r$ = 1.8 and
those with $r > 2.5$ are that, as expected, $M_{sub}$ decreases
and $M_{nb}$ increases with decreasing $r$.
The increase in $M_{nb}$ implies an increase
in $\mu_{IMF}$. 
Alternatively the changes in the C/O, $\Delta Y/\Delta {\rm O}$, and $Z$/O
values are negligible.
These models would imply that the $M_{nb}$ determination by
Walter et al. (1997) is not correct.

\subsubsection { O-rich Outflow Models with $M_{nb}$ =  0 }

We have produced O-rich outflow models ($\gamma \ne 0.00$)
 for II Zw 33 with $r$ = 1.8 and $M_{nb}$ = 0.0
(see Tables 5 and 6). 
The O-rich outflow models predict C/O, $\Delta Y/\Delta {\rm O}$, and $Z$/O values
higher than the closed-box models (see Table 6).  The C/O and $Z$/O values
are not known for II Zw 33; our models predict that if O-rich outflows have
been important during the evolution of this object its C/O and $Z$/O
values should be considerably higher than those observed in
other irregular galaxies (see \S 7.2).

\begin{planotable}{ccccccc}
\tablecolumns{6}
\tablewidth{0pt}
\tablecaption{Abundance Ratios for II Zw 33 }
\tablehead{\colhead{$t_g$(Gyr)} &  \colhead{$r$} 
& \colhead{$\gamma$} &
\colhead{C/O} & \colhead{$\Delta Y/\Delta {\rm  O}$} &\colhead{$Z$/O}}
\startdata
0.1  & 2.502 & 0.00 & 0.169 & 3.011 & 1.884 \nl
     & 1.800 & 0.00 & 0.168 & 2.890 & 1.852 \nl
     & 1.800 & 0.29 & 0.170 & 3.204 & 1.939 \nl
\tablevspace{1.5ex}
1.0  & 2.632 & 0.00 & 0.311 & 3.789 & 2.317 \nl
     & 1.800 & 0.00 & 0.294 & 3.695 & 2.273 \nl
     & 1.800 & 0.33 & 0.383 & 4.419 & 2.592 \nl
\tablevspace{1.5ex}
10.0 & 2.750 & 0.00 & 0.328 & 4.247 & 2.447 \nl
     & 1.800 & 0.00 & 0.325 & 4.173 & 2.429 \nl
     & 1.800 & 0.38 & 0.425 & 5.284 & 2.856 \nl
\tablevspace{1.5ex}
\enddata
\end{planotable}

\section {A TYPICAL IRREGULAR GALAXY}

For NGC 1560 and II Zw 33 we have only $\mu_{IMF}$, $M_{nb}$, and O as
observational constraints; these constraints are not enough to decide
if O-rich outflows have been present in these objects. We have decided to
model a typical irregular galaxy because for it we can use average observational
values 
for  $\mu$, O, C/O, $\Delta Y/\Delta{\rm O}$, and $Z$/O,
derived {from} a set of well observed irregular galaxies; 
 these observational constraints will permit us to address
the issue of the importance of O-rich outflows for the evolution of
the typical dwarf irregular galaxy.

\begin{planotable}{lccccc}
\tablecolumns{6}
\tablewidth{0pt}
\tablecaption{Properties of Selected Galaxies \label{tab:galaxies}}
\tablehead{\colhead{Galaxy} & \colhead{$\log (M_{total}/\msun)$} &
\colhead{$\log (M_{gas}/\msun)$} & \colhead{$\log \mu$} & \colhead{$Y$} & 
\colhead{10$^3$O\tablenotemark{a}}  }
\startdata
I Zw 18  &  8.26\phm{100} & 8.21 & $-0.05$ & 0.230 & 0.317 \nl
UGC 4483 &  8.07\phm{100} & 7.95 & $-0.12$ & 0.239 & 0.639 \nl
Mrk 600  &  8.88\phm{100} & 8.76 & $-0.12$ & 0.240 & 1.967 \nl
SMC      &  9.12\phm{100} & 8.71 & $-0.41$ & 0.237 & 2.268 \nl
II Zw 40\tablenotemark{b}& 9.46\phm{100} & 8.62 & $-0.84$ & 0.251 & 2.672 \nl
IC 10    & 10.03\phm{100} & 9.74 & $-0.29$ & 0.240 & 2.775 \nl
II Zw 70 &  9.14\phm{100} & 8.67 & $-0.47$ & 0.250 & 3.287 \nl
NGC 6822 &  9.23\phm{100} & 8.31 & $-0.92$ & 0.246 & 3.305 \nl
LMC      &  9.78\phm{100} & 8.88 & $-0.90$ & 0.250 & 4.133 \nl
NGC 4449 & 10.91\phm{100} & 9.63 & $-1.28$ & 0.251 & 4.526 \nl
average  &  9.29\phm{100} & 8.75 & $-0.54$ & 0.243 & 2.589 \nl 
\enddata
\tablenotetext{a}{The O gaseous values have been multiplied by 1.58 (0.2 dex, see text).}
\tablenotetext{b}{Values quoted for the northern cloud. Molecular and ionized
hydrogen is added.}
\end{planotable}

In what follows we will estimate the general properties of a typical
irregular galaxy. We will use the same set of galaxies
that was used by CCPS. These galaxies were chosen because their properties
are well known, in particular the chemical composition of their
gaseous content.

For the irregular galaxies in the CCPS sample we do not have a $\mu_{IMF}$
value because we do not know which is the contribution of the halo dark
matter, $M_{nb}$, to $M_{total}$. Therefore the observed $\mu$ value is a
lower limit to $\mu_{IMF}$  (see equation 4).

\subsection{The $\mu$ Value}

The $\mu$ value depends on the $M_{gas}$ to $M_{total}$ ratio. 
We will revise the $M_{gas}$
and $M_{total}$ values adopted by CCPS for each galaxy and the new adopted values
will be presented in Table 7.

CCPS neglected the contribution of ${\rm H}_2$ to $M_{gas}$ due to the low CO
content of the irregular galaxies. Nevertheless, there are observational
and theoretical
considerations that favor a
CO-to-${\rm H}_2$ conversion factor, $X_{\rm CO-H_2}$,
that increases for systems of lower metallicity (Maloney \& Black 1988;
Wilson 1995; Arimoto, Sofue, \& Tsujimoto 1996). Therefore we can not
exclude the possibility of having low-metallicity irregular galaxies
with a relatively high molecular hydrogen content. 

The ${\rm H}_2$ mass estimated for
II Zw 40 by Tacconi \& Young (1987), for NGC 6822 by Israel (1997),
and adopted for II Zw 33 by Walter et al. (1997) is about 10 \% of H I mass.
Consequently, we will
multiply the H I + He I gaseous mass of irregular galaxies by a factor
of 1.1 to take into account the contribution due to  ${\rm H}_2$, with the
exception of I Zw 18, and UGC 4483, where no such correction will
be applied (Carigi \& Peimbert 1998).

Madden et al. (1997)
recently estimated an unusually high ${\rm H}_2$
column density (a factor
of five that of H I) in three regions of IC 10,
their result is based
on [C II] 158 micron observations and an argument of thermal balance.
If this result is applied to the whole galaxy one finds that $\mu=1.2$ 
which is impossible since by definition $\mu$ has to be smaller than 1.
We consider that there is no room for such high amounts of ${\rm H}_2$ in dwarf
irregular galaxies (see also Lequeux 1996).

To produce a homogeneous set of $\mu$ values we need to estimate $M_{total}$
at a given distance {from} the galactic center. Wherever possible we
have chosen the Holmberg radius, $R_H$, because most of the gas and stars are
inside it, for larger radii $M_{nb}$ becomes larger and $\mu$ becomes smaller
deviating more {from} $\mu_{IMF}$.

Based in part on the previous discussion we have introduced the
following changes to Table 1 of CCPS and have generated the $M_{total}$ 
and $M_{gas}$
values of Table 7:
a) we have added a 10 \% to the H I + He I gaseous mass to consider
the H$_2$ contribution for all galaxies except 
for II Zw 40, where H$_2$ is explicitly taken by Tacconi \& Young (1987),
and I Zw 18 and UGC 4483,
 where we are assuming a null H$_2$; 
b) we have adopted a Hubble constant of 
$H_0 = 100 h$ kms Mpc$^{-1}$ with $h=0.65$,
while CCPS adopted $h = 0.70$, errors in the distance, $d$, will
alter $\mu_{IMF}$ because $M_{gas} \propto d^2$ while $M_{total} \propto d$;
c) we have revised for each galaxy the determination of $M_{total}$.
The discussion on the $M_{total}$ and $\mu$ determinations for each galaxy
follow.

I Zw 18.--- The total mass within $R_H$ (Staveley-Smith, Davies, \& Kinman 1992, hereinafter SDK)
give a ratio of $M_{gas}$ to $M_{total}$ greater than one. On the
other hand, assuming that the high radial velocity gradient is 
due to rotation, Petrosian et al. (1997) derive a total
mass (within a radius of 0.48 kpc) higher than that derived by
SDK. The difference is due to the higher rotational 
velocity estimated by Petrosian et al.. Our $M_{total}$ and $M_{gas}$ values are 
those adopted by Carigi \& Peimbert (1998), 
which are a compromise between the values
given by SDK and Petrosian et al.. 
Incidentally, the $\mu$ value adopted here
is close to that adopted by CCPS.

UGC 4483.--- There are two determinations of $M_{total}$, one by SDK
and another by
Lo, Sargent, \& Young (1993),
which together with the $M_{gas}$ derived by SDK yield an average
$\mu$ value of 0.76.
The radius at which $M_{total}$ is derived, 1.42 kpc (putting UGC 4483
at the distance  given by SDK but with $h=0.65$), 
is close to its $R_H =$ 1.28 kpc (SDK). 

Mrk 600.--- We adopted as the total mass within $R_H$ that derived by SDK
but assuming an $h = 0.65$.

SMC, LMC, \& II Zw 40.--- These three galaxies 
have a total mass derived within a radius that is lower than 
$R_H$, being II Zw 40 the extreme case. 
This is not a problem as long as $\mu$ does not suffer a
significant change when going {from} this radius to $R_H$.
The radius at which the total mass of SMC is derived (Hindman 1967)
is close to its $R_H$ (Balkowski, Chamaraux, \& Weliachew 1978), therefore we would not have expected 
a very different $\mu$ value if $M_{total}$ had
been determined inside $R_H$ (we are also assuming that
the contribution {from} the gas beyond $R_H$ is
negligible).  As the distance determined by Welch et al. (1987)
of 61 kpc has now been used, as opposed to 70 kpc, its $M_{total}$ and $M_{gas}$
values have changed accordingly. On the other hand, according to 
Kunkel, Demers, \& Irwin (1996) most of the mass of the LMC is located 
within the inner 5 $\deg$, we are thus not making a big 
mistake by using the $\mu$ associated to the inner region of 4.2 
$\deg$ (Lequeux et al. 1979); apparently there is no
massive dark halo in LMC (Kunkel et al. 1997). The H I core of II Zw 40 was studied by
Gottesman \& Weliachew (1972) and its corresponding $\mu$ was adopted by CCPS {from}
Lequeux et al.. {From} the $M_{total}$ values derived by Brinks \& Klein (1988),
we find $\mu$ value of 0.14  for the northern  cloud of 
II Zw 40 (see Table 7).
This low value of $\mu$ might imply the presence of a significant amount of dark matter.

IC 10.--- The Keplerian estimate of $M_{total}$ (Shostak 1974)
was derived within a radius which is very close to 
$R_H$ ($R_H$=4.0 kpc, if the distance to the galaxy is taken to be
3 Mpc). Recent determinations of the distance to IC 10 put it
close to 1 Mpc (e.g., Wilson et al. 1996). This value contrasts with the 3 Mpc 
adopted by Lequeux et al. {from} Sandage \& Tammann (1975). We have adopted 1.5 Mpc and at 
this distance $\mu = 0.51$.

II Zw 70 \& NGC 6822.--- The total masses of these two galaxies are
derived within a radius which is greater than their 
$R_H$ values by about a factor of two (Balkowski et al. 1978;
 Gottesman \& Weliachew 1977). 
By using these $M_{total}$ values,
we may be underestimating the value of $\mu$ as compared with
the values derived for other galaxies in the sample. A very detailed
modeling of the rotation curve of these two galaxies is needed to know
the contribution of a dark halo to the total mass.

NGC 4449.--- The total mass of this galaxy, within a radius of 37 kpc
(by far greater than its optical radius)
has been estimated recently by Bajaja, Huchtmeier, \& Klein (1994). Its $\mu = 0.052$ is considerably
lower than those usually estimated for dwarf irregular galaxies, in particular,
lower than the other galaxies of our sample. The rotation curve indicates the
presence of dark matter in its extended H I halo but, as in the case of NGC 6822
and II Zw 70, a very detailed modeling of its rotation curve is still needed.

It is interesting to note that the mean $\mu$ value presented in Table 7
differs only by 2 \% {from} the mean $\mu$ value obtained by CCPS.

\subsection{ O, C/O, $\Delta Y/\Delta {\rm O}$, and $Z$/O }

Columns 5 and 6 of Table 7 present the helium  and the oxygen
abundances by mass, the data are the same as those presented by CCPS.
In the last two lines of Tables 9 and 11 we present the C/O, 
$\Delta Y/\Delta {\rm O}$, and $Z$/O values derived by CCPS
{from} their sample of irregular galaxies.

More recent determinations of He abundances
permit to derive other $\Delta Y/\Delta {\rm O}$ values. {From} the data of 
Izotov, Thuan, \& Lipovetski (1997)
on extragalactic H II regions it is obtained that 
$\Delta Y/\Delta {\rm O} = 3.1 \pm 1.4$.
Based on observations by many authors Olive, Steigman, \& Skillman (1997) obtain a
pregalactic helium abundance, $Y_p$, of 0.234 $\pm$ 0.002, alternatively
Izotov et al. find $Y_p = 0.243 \pm  0.003$. By adopting 
$Y_p = 0.240 \pm  0.006$
and combining this value  with the $Y$ and O abundances of the
galactic H II region M17,  that amount to 0.280 $\pm$ 0.006 and
$(8.69 \pm 1.3) \times 10^{-3}$, respectively
(Peimbert, Torres-Peimbert, \& Ruiz 1992),
it follows that $\Delta Y/\Delta {\rm O} = 4.6 \pm 1.1$.  
We have added 0.08 dex to the gaseous
O abundance to consider the fraction of O atoms embedded in dust (Esteban et al.
1998) . Finally, {from} fine structure in the main sequence based on Hipparcos
parallaxes Pagel \& Portinari (1998) obtain that 
$\Delta Y/\Delta {\rm O} = 5.6 \pm 3.6$. These
three $\Delta Y/\Delta {\rm O}$ values are in good agreement with the value 
presented in Tables 9 and 11.
{\pspicture(0.5,-1.5)(12.0,11.3)
\rput[tl]{0}(-0.5,11.5){\epsfxsize=10cm
\epsffile{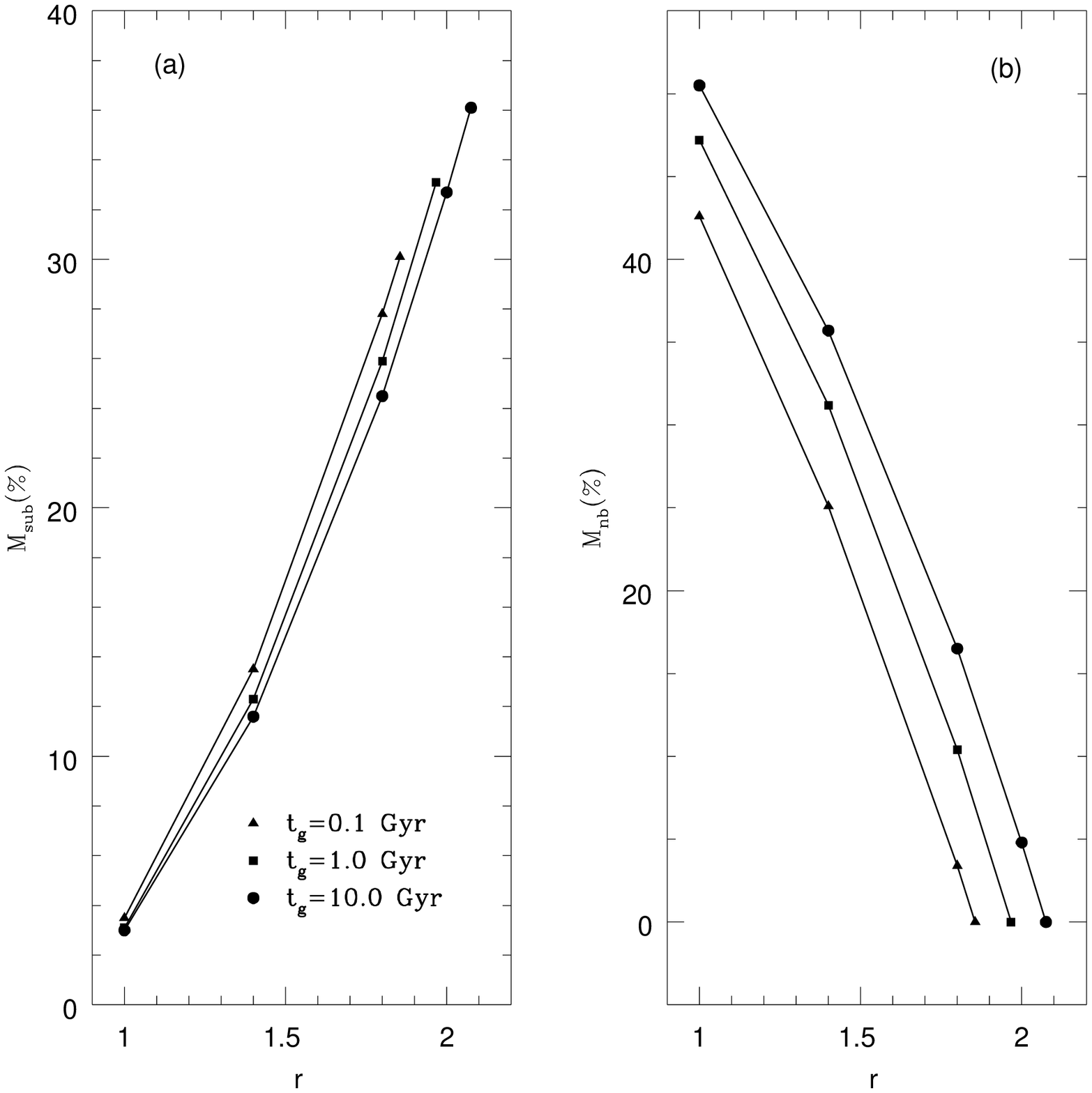}}
\rput[tl]{0}(0.4,1.5){
\begin{minipage}{8.3cm}
  \small\parindent=3.5mm {\sc Fig.}~1.--- 
 Mass fractions of (a) substellar objects, $M_{sub}$ ,
 and of (b) non baryonic dark matter, $M_{nb}$,
 for different $r(0.01,85)$ values.
 The data correspond to closed-box models with continuous SFRs for
 different ages of the typical irregular galaxy (see Table  8).
\end{minipage}}
\endpspicture}

\subsection{Closed-box Models with Continuous SFRs }

The properties of the typical irregular galaxy were obtained {from} the
galaxies presented in Table 7 and are given in the last row of
this table.

We have computed  closed-box models with continuous SFRs and different ages for the 
typical irregular galaxy. 
All the models
reproduce the observational constraints, $\mu = 0.288$ and O = $2.589 \times 10^{-3}$.
For the typical irregular galaxy we do not know the amount of $M_{nb}$, therefore
the observed $\mu$ value is a lower limit to $\mu_{IMF}$. Consequently, we have
computed models for a range of $r(0.01,85)$ values; we think it is unlikely that the $r$
 value
is smaller than one (see Table 1), and $r_{max}$ is the value for 
$M_{nb} = 0.0$,
$r$ can not be higher than $r_{max}$ because $M_{nb}$ would become negative.
In Table 8 we present  the model results for the different mass fractions
defined in this paper.

The distributions of $M_{sub}$ and $M_{nb}$ for different $r(0.01,85)$
values are plotted in Figure 1. {From} this figure it can
be seen that $M_{sub}$ increases with $r$ and $M_{nb}$ decreases with $r$.
The increase of $M_{sub}$ with $r$ is due to an increase of the slope of
the low-mass end of the IMF with $r$.
The decrease of $M_{nb}$  with $r$ is due to the lower efficiency
in the O production and therefore to the decrease of $\mu_{IMF}$
(see Table 8 and equation 4).
Furthermore for a given $r$ value $M_{sub}$ decreases with the age of 
the model while $M_{nb}$ increases with it.

\begin{planotable}{cccccccc}
\tablecolumns{8}
\tablewidth{0pt}
\tablecaption{Continuous SFR Models for the Typical Galaxy
\tablenotemark{a}}
\tablehead{\colhead{$t_g$(Gyr)} &  \colhead{$r$} & 
\colhead{$M_{sub}(\%)$} & \colhead{$M_{vl}(\%)$} & \colhead{$M_{rest}(\%)$}
& \colhead{$M_{rem}(\%)$} &\colhead{$M_{nb}(\%)$} & \colhead{$\mu_{IMF}$}}
\startdata
0.1  & 1.000 &  3.5 &  9.1 & 14.8 & 0.3 & 42.6 & 0.52 \nl
     & 1.400 & 13.5 & 14.3 & 17.0 & 0.4 & 25.1 & 0.40 \nl
     & 1.800 & 27.8 & 19.6 & 19.1 & 0.4 &  3.4 & 0.31 \nl
     & 1.855 & 30.1 & 20.3 & 19.4 & 0.5 &  0.0 & 0.30 \nl
\tablevspace{1.5ex}
1.0  & 1.000 &  3.1 &  8.2 & 10.6 & 1.2 & 47.2 & 0.56 \nl
     & 1.400 & 12.3 & 13.3 & 12.1 & 1.4 & 31.2 & 0.43 \nl
     & 1.800 & 25.9 & 18.3 & 14.1 & 1.6 & 10.4 & 0.33 \nl
     & 1.967 & 33.1 & 20.6 & 14.8 & 1.8 &  0.0 & 0.30 \nl
\tablevspace{1.5ex}
10.0 & 1.000 &  3.0 &  7.8 &  6.0 & 3.0 & 50.5 & 0.60 \nl
     & 1.400 & 11.6 & 12.5 &  7.0 & 3.5 & 35.7 & 0.46 \nl
     & 1.800 & 24.5 & 17.3 &  7.8 & 4.2 & 16.5 & 0.36 \nl
     & 2.000 & 32.7 & 19.9 &  8.4 & 4.5 &  4.8 & 0.31 \nl
     & 2.075 & 36.1 & 20.9 &  8.6 & 4.7 &  0.0 & 0.30 \nl
\enddata
\tablenotetext{a}{$M_{sub} + M_{vl} + M_{rest} 
+ M_{rem} + M_{gas} + M_{nb} = 100.0\%$, with $M_{gas} = 29.7\%$}
\end{planotable}

In Table 9 we present the predicted abundance ratios by our closed-box
models with continuous SFRs, the predicted abundance ratios by CCPS, and the observed ratios.
{From} this table it can be seen that:
a) the predicted abundance ratios for a given model age are almost independent 
of $r$; 
b) the predicted ratios increase with model age due to the production of
$Y$ and C  by intermediate mass stars, while O is produced only by
massive stars;
c) the C/O and $Z$/O  values point to models with 
ages not older than one Gyr, while the 
$\Delta Y/\Delta {\rm O }$ value indicates older ages.
d) our models predict considerably larger values for 
C/O, $\Delta Y/\Delta {\rm O }$, and $Z$/O than the models by CCPS,
this result is due to the difference between the yields by WLW \& WW
and those by Maeder (1992).

\begin{planotable}{ccccccc}
\tablecolumns{5}
\tablewidth{0pt}
\tablecaption{Abundance Ratios for the Typical Galaxy (Continuous SFRs)}
\tablehead{\colhead{$t_g$(Gyr)} &  \colhead{$r$} & 
\colhead{C/O} & \colhead{$\Delta Y/\Delta {\rm  O}$} &\colhead{$Z$/O}}
\startdata
0.1  & 1.000 & 0.165 & 2.788 & 1.832  \nl
     & 1.400 & 0.166 & 2.873 & 1.855 \nl
     & 1.800 & 0.167 & 2.965 & 1.876 \nl
     & 1.855 & 0.167 & 2.978 & 1.883 \nl
     & 3.240\tablenotemark{a} 
             & 0.135 & 1.783 & 1.736 \nl  
\tablevspace{1.5ex}
1.0  & 1.000 & 0.283 & 3.608 & 2.247 \nl
     & 1.400 & 0.292 & 3.663 & 2.273 \nl
     & 1.800 & 0.303 & 3.721 & 2.299 \nl
     & 1.967 & 0.307 & 3.747 & 2.315 \nl
     & 3.290\tablenotemark{a} 
             & 0.225 & 2.597 & 1.927 \nl  
\tablevspace{1.5ex}
10.0 & 1.000 & 0.320 & 4.093 & 2.415 \nl
     & 1.400 & 0.322 & 4.130 & 2.427 \nl
     & 1.800 & 0.324 & 4.170 & 2.439 \nl
     & 2.000 & 0.324 & 4.191 & 2.439 \nl
     & 2.075 & 0.325 & 4.199 & 2.445 \nl
     & 3.390\tablenotemark{a} 
             & 0.240 & 2.946 & 1.938 \nl
\tablevspace{1.5ex}
Obs\tablenotemark{b}           & & 0.212 & 4.48 & 1.85 \nl
errors($\pm$)\tablenotemark{b} & & 0.071 & 1.02 & 0.20 \nl
\enddata
\tablenotetext{a}{These lines corresponds to closed-box models {from} CCPS}
\tablenotetext{b}{Taken {from} CCPS}
\end{planotable}

Our sample in Table 7 was chosen to have a large spread in $Y$ and O to
derive a meaningful $\Delta Y/\Delta {\rm O }$ value, but our sample is not
completely homogeneous because the O abundances for I Zw 18 and UGC 4483
are very small and the $M_{total}$ for NGC 4449 was estimated {from}
observations at distances far away {from} $R_H$. To have a more homogeneous
sample we can eliminate the three objects in Table 7 that deviate most
{from} the average $\mu$ and O values: I Zw 18, UGC 4483, and NGC 4449.
For this reduced sample we obtain the following average quantities:
log$\mu=-0.56$ and $10^3 {\rm O}=2.915$,
in very good agreement with the values adopted for the typical irregular
galaxy. Moreover, the chemical evolution models computed to adjust the
average values of the reduced sample are very similar to those computed
for the typical irregular galaxy.

\subsection{ Closed-box Models with Bursting SFRs}

The presence of old populations in many of the dwarf irregular
galaxies in the local group indicates that star formation
started $\sim 10 $ Gyr ago (e.g., Mateo 1998; Pagel \& Tautvai\v{s}ien\.e 1998).
The nature of the star formation histories in these galaxies is diverse
(e.g., Mateo 1998). The most massive dwarf irregular galaxies appear to have
had a continuous SFR, while ordinary dwarf irregular galaxies appear to have 
undergone
a gasping SFRs (e.g., Tosi 1998). It is interesting then to see if by
changing the continuous nature of the SFR to a discontinuous one (bursting),
our results presented in \S 6.3 change drastically.

Figure 2 shows in six panels the evolution of: the SFR,
$\mu_{IMF}$, O,  C/O, $\Delta Y/\Delta 
{\rm O }$, and $Z$/O. The model is for $t_g= 1.0 $ Gyr  and has three bursts
each lasting 40 Myr. The behavior of O, 
C/O,  $\Delta Y/\Delta {\rm O }$, and $Z$/O  are as expected. The oxygen
abundance increases suddenly during each burst and then stays constant in the
quiescent phase. On the other hand, C, $Y$ and $Z$ increase during the
quiescent phase because they are also produced by intermediate mass stars.
There is a period of time after the second or third burst in which these ratios
decrease because the growth rate of the O abundance is higher than those
of C, $Y$, and $Z$. 
The spike in the C/O,  $\Delta Y/\Delta {\rm O }$, and $Z$/O values
at the beginning of the model is due to the C and $Y$ production
in very massive stars relative to that of O.

Table 10 is similar to Table 8 but with an extra column that gives the
number of bursts, $n$, of each model. 
We did not compute a bursting SFR model for 0.1 Gyr, because the 0.1-Gyr
model with continuous SFR presented in Tables 8 and 9 can be considered
as a bursting model with single burst lasting 100 Myr.
It can be
seen {from} Table 10 that the $r$ parameter is not very sensitive to $n$. The
$r$ parameter increases slightly as we reduce the number of bursts.

\begin{planotable}{ccccccccc}
\tablecolumns{9}
\tablewidth{0pt}
\tablecaption{Bursting SFR Models for the Typical Galaxy
\tablenotemark{a}}
\tablehead{\colhead{$t_g$(Gyr)} & \colhead{$n$} & \colhead{$r$} &
\colhead{$M_{sub}(\%)$} & \colhead{$M_{vl}(\%)$} & \colhead{$M_{rest}(\%)$}
& \colhead{$M_{rem}(\%)$} &\colhead{$M_{nb}(\%)$} & \colhead{$\mu_{IMF}$}}
\startdata
1.0  & 2 & 1.930 & 27.5 & 17.6 & 13.2 & 1.6 & 10.4 & 0.33 \nl
     & 3 & 1.894 & 27.2 & 17.8 & 13.2 & 1.7 & 10.4 & 0.33 \nl
     & 5 & 1.860 & 26.8 & 18.0 & 13.4 & 1.7 & 10.4 & 0.33 \nl
     & $\infty$\tablenotemark{b} 
         & 1.800 & 25.9 & 18.3 & 14.1 & 1.6 & 10.4 & 0.33 \nl
\tablevspace{1.5ex}
10.0 & 2 & 1.841 & 24.0 & 16.4 &  9.0 & 4.4 & 16.5 & 0.36 \nl 
     & 3 & 1.832 & 24.2 & 16.7 &  8.4 & 4.5 & 16.5 & 0.36 \nl
     & 5 & 1.818 & 24.1 & 16.7 &  8.8 & 4.2 & 16.5 & 0.36 \nl
     & $\infty$\tablenotemark{b} 
         & 1.800 & 24.5 & 17.3 &  7.8 & 4.2 & 16.5 & 0.36 \nl
\enddata
\tablenotetext{a}{$M_{sub} + M_{vl} + M_{rest}
+ M_{rem} + M_{gas} + M_{nb} = 100.0\%$, with $M_{gas} = 29.7\%$}
\tablenotetext{b}{These lines corresponds to continuous SFRs}
\end{planotable}

The abundance ratios for the typical irregular galaxy with  bursting SFRs
are given in Table 11. This Table is similar to Table 9 except for the inclusion
of a new column representing the number of bursts of the model and the absence
of the 0.1-Gyr model. All of the remarks drawn {from} Table 9 can also
be drawn {from} Table 11; in particular, 
the 1.0-Gyr and 10-Gyr models reproduce well the observed 
$\Delta Y/\Delta {\rm O }$
value but overestimate the observed C/O and $Z$/O values.
A slight improvement in the comparison of the predicted versus observed
C/O and $Z$/O
values is introduced by  bursting SFRs (specially when $n =2$),
but high C/O and $Z$/O values continue to be predicted.
Furthermore, {from} Figure 2 and the observed values in Table 11,
it can be seen that after a burst the predicted versus the observed
C/O and $Z$/O values become closer, but the $\Delta Y/\Delta {\rm O}$
values become farther apart.

\begin{planotable}{cccccccc}
\tablecolumns{6}
\tablewidth{0pt}
\tablecaption{Abundance Ratios for the Typical Galaxy (Bursting SFRs) }
\tablehead{\colhead{$t_g$(Gyr)} & $n$ &  \colhead{$r$} &
\colhead{C/O} & \colhead{$\Delta Y/\Delta {\rm  O}$} &\colhead{$Z$/O}}
\startdata
1.0  & 2 & 1.930 & 0.289 & 4.024 & 2.349 \nl
     & 3 & 1.894 & 0.304 & 3.943 & 2.349 \nl
     & 5 & 1.860 & 0.319 & 3.894 & 2.358 \nl
     & $\infty$\tablenotemark{a} 
         & 1.800 & 0.303 & 3.721 & 2.299 \nl
\tablevspace{1.5ex}
10.0 & 2 & 1.841 & 0.291 & 4.355 & 2.389 \nl
     & 3 & 1.832 & 0.303 & 4.288 & 2.406 \nl
     & 5 & 1.818 & 0.309 & 4.171 & 2.408 \nl
     & $\infty$\tablenotemark{a} 
         & 1.800 & 0.324 & 4.170 & 2.439 \nl
\tablevspace{1.5ex}
Obs\tablenotemark{b}           & & & 0.212 & 4.48 & 1.85 \nl
errors($\pm$)\tablenotemark{b} & & & 0.071 & 1.02 & 0.20 \nl
\enddata
\tablenotetext{a}{These lines corresponds to continuous SFRs}
\tablenotetext{b}{Taken {from} CCPS}
\end{planotable}

\section{DISCUSSION}

\subsection{The $r$ Value}

 The $r$ value is almost independent of moderate changes in the high-mass 
limit  of the IMF,
alternatively it depends strongly on the low-mass limit and on the
slope of the IMF at the low-mass end
(see Table 1).

The average $r(0.01,85)$ value for the globular clusters NGC 6752 and NGC 7099
amounts to 1.85, while for $r(0.08,85)$ amounts
to 1.32 (see Table 1). 
These $r$ values are
higher than those for the solar vicinity and might mean two different things:
a) that the $r$ value for the solar vicinity is not well known and that
the $r$ value for globular clusters is representative of the solar
vicinity, or b) that systems with lower metallicity have higher $r$ values.
In this discussion we have not considered NGC 6397 due to the possible
selective loss of low-mass stars by evaporation and tidal encounters.

A system with $r>1$ has a larger mass fraction of objects with $m<0.5$
than a system with $r=1$, this is true for any value of $m_l<0.5$,
including the case when $M_{sub}=0.0$

\subsection{Model Results}

We have computed closed-box models for NGC 1560 and II Zw 33 that match 
the observed O and $\mu_{IMF}$ values. For the 10 Gyr models the $r$
values are equal to 1.73 and 2.75, respectively.

If the IMF is universal it follows that the same $r$ should apply to all
galaxies. {From} globular clusters it is obtained that $r \sim 1.8$,
this is the best $r$ determination available and it might apply to all
galaxies.

Since closed-box models with continuous SFRs and $M_{nb}=0.0$ yield
$2.5 < r < 2.75$ for II Zw 33, we decided to compute other types
of models with $r=1.8$ for this galaxy. The closed-box continuous SFR
models with $r=1.8$ require that 28 \% $< M_{nb} <$ 37 \%, in
contradiction with the results of $M_{nb}=0.0$ derived by Walter
et al. (1997). O-rich continuous SFR models with $r=1.8$ and $M_{nb}=0.0$
imply C/O $>$ 0.383 for models older than 1.0 Gyr. This C/O value is higher 
than the highest value detected for an irregular galaxy (0.248 for
30 Doradus in LMC) and considerably higher than 0.158, the average C/O value
for the eight irregular galaxies for which C/O is known (Garnett et al.
1995, 1997, Kobulnicky et al. 1997).
The C/O values for II Zw 33 is not known and conceivable could be
considerably higher than those of other irregular galaxies but it seems
unlikely.

We have also computed closed-box models with continuous SFRs
for the typical irregular galaxy that 
 match the observed $\mu$, O, and $\Delta Y/\Delta {\rm O}$ values
for different $r$ values (see Tables 8 and 9). To choose one of these models we need
to know the $r$ or the $M_{nb}$ value.
Based on the models for NGC 1560 and on the $r$ values for
globular clusters we consider that the best models should be around $r=1.8$.

A maximun value of $r=2.1$ is obtained  when $M_{nb} \to 0.0$
for the typical irregular galaxy (see Table 8).

 We do not know which is the behavior of the IMF for $m<0.1$, nor the
value of $r$, but by assuming that $\alpha_1$ is the same down to $m=0.01$,
the solution for the typical irregular galaxy with $r(0.01,85)=1.80$ 
at 10 Gyr implies that
$M_{sub}=24.5$ \% and $M_{nb}=16.5$ \%.
If $M_{sub}$ is smaller than that obtained {from} a given  $\alpha_1$,
 it is possible to obtain a
closed-box model that fits the same observational constraints by increasing
$M_{nb}$ (see Table 8).

We consider that O-rich outflows are not very important for the typical
irregular galaxy because O-rich
outflow models predict higher C/O and Z/O 
values than those observed.

The changes in the final abundance ratios between the 0.1-Gyr and the
1.0-Gyr models (see Table 9) are more important than the changes 
between different SFR histories, for a given $t_g$ (see Table 11).
For 1.0 Gyr $\leq t_g \leq$ 10 Gyr the changes between bursting
and continuous SFR models are very small (see Table 11).
For $t_g=0.1$ Gyr, the continuous SFR model can be considered as a
single-burst model and the differences between this model and those for
1.0 Gyr and 10 Gyr are very significant (see Table 9).

We also consider that outflow of well-mixed material is not important.
Outflow models of well-mixed material will have lower $r$ and
$M_{sub}$ values than those presented in Tables 4, 5, 8, and 10; but to produce
drastic reductions in $r$ these models require the ejection of large
amounts of gas to the intergalactic medium, that have not yet been
observed around irregular galaxies.
The previous discussion is based on CCPS models for outflow of well-mixed
material; their models were made for $r=1$ while their closed-box
model for 10 Gyr gives $r=3.39$, and the ratio of the ejected mass
to the mass left in the galaxy is 7.95.

Furthermore, in general infall of material with pregalactic abundances, $Y=Y_p$ and $Z=0$,
is not important because these models are not able to 
match low O values with moderately low $\mu_{IMF}$ values
(Peimbert et al. 1994).

\subsection { CCPS O-rich Outflow Models}

Are the closed-box models presented in Tables 8-11 the only possibility to
adjust the observational constraints of the typical irregular galaxy? The answer is no.
By using the yields by Maeder (1992)
CCPS have shown that
 it is also possible to adjust the
observational constraints by means of O-rich outflow models.

The two reasons given by CCPS to support O-rich outflow models
were the high $r$ and the low  $\Delta Y/\Delta {\rm O}$ values
predicted by the closed-box models; 
for the 10 Gyr model these values amount to 3.390 and 2.946,
respectively (see Table 9).
This CCPS model was made under the $M_{nb}=0.0$ assumption.
Our 10-Gyr closed-box model for the typical irregular galaxy
with $M_{nb}=0.0$ yields $r=2.075$,
and $\Delta Y/\Delta {\rm O}=4.199$.
The differences between our model and the CCPS model are only due to
differences in the adopted yields.

By introducing $M_{nb}$ values different {from} zero in the CCPS models 
it is possible to reduce $r$ to a reasonable value for a closed-box
model; but  $\Delta Y/\Delta {\rm O}$ 
would still be lower than observed because it is almost independent
of $r$.
\begin{figure*}[ht]
\pspicture(0.5,-1.5)(15.0,19.0)
\rput[tl]{0}(1.0,18.0){\epsfxsize=18cm
\epsffile{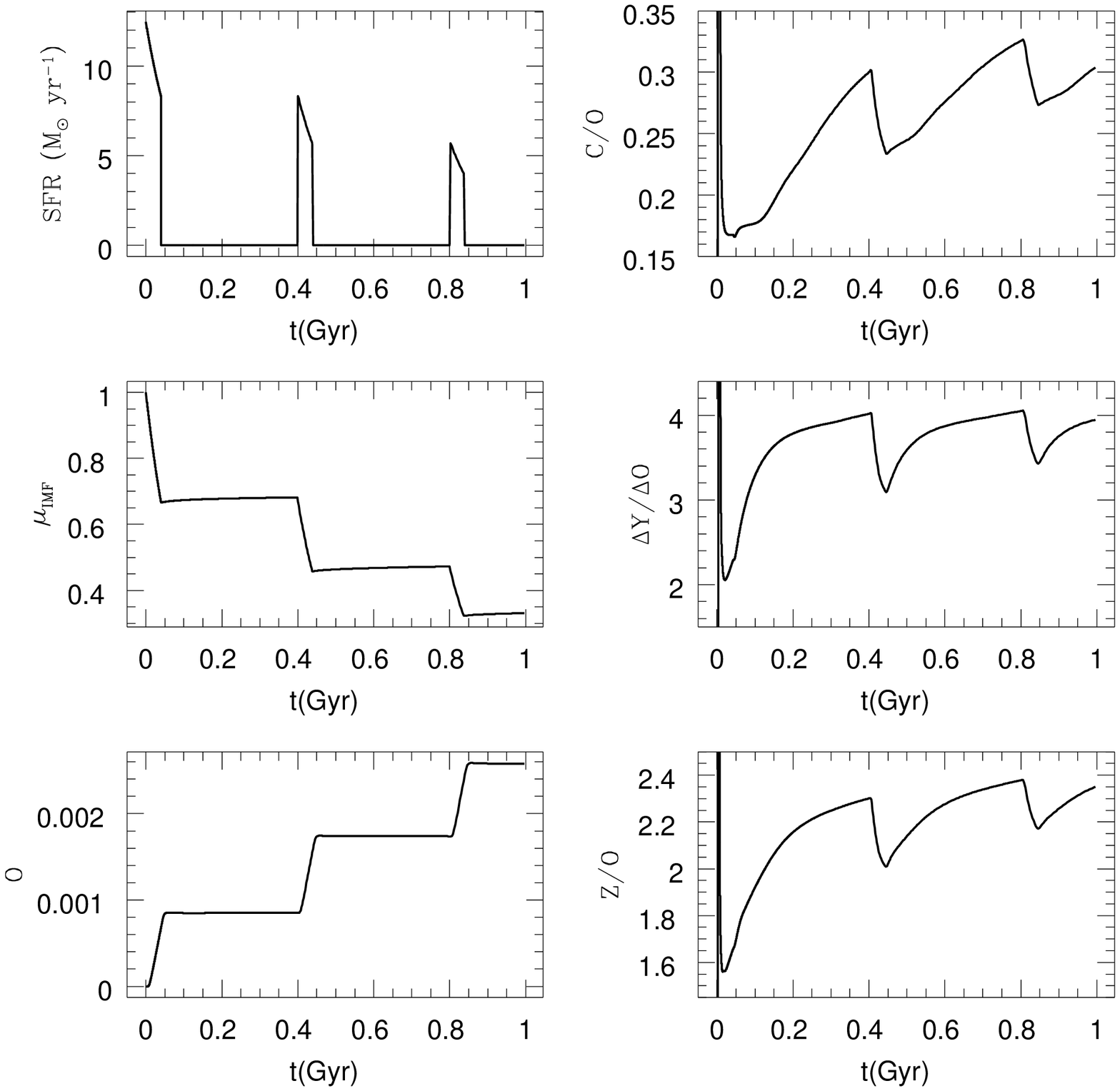}}
\rput[tl]{0}(0.5,0.0){
\begin{minipage}{18cm}
  \small\parindent=3.5mm {\sc Fig.}~2.---
Bursting SFR model for the typical irregular galaxy with $t_g=1.0$ Gyr,
$n=3$, and $r(0.01,85)=1.894$ (see Table 11). C, O, $Y$, and $Z$ are given
by mass.
\end{minipage}}
\endpspicture
\end{figure*}

\section{CONCLUSIONS}

The IMF {from} globular clusters 
has a larger fraction of low-mass stars, and consequently 
 a larger value of $r$, than the KTG IMF.
This result is based on the assumption that the IMF slopes for $m > 0.5$
are the same for all objects. The difference in the $r$ value, if real,
could be due to the lower metallicity of the globular clusters relative to the
solar vicinity. Alternatively, considering that the IMF seems to be metallicity
independent at higher masses, the difference in the $r$ value could be
due to observational errors; since the accuracy of the IMF determination
for globular clusters is higher than for the solar vicinity maybe the
lower end of the KTG IMF should be modified to agree with that derived
{from} globular clusters.

The $r$ values required by the closed-box models based on the yields by WLW \& WW
for  NGC 1560 and II Zw 33, where $M_{nb}$ has been determined,
are considerably higher than one. Moreover, the $r$ value for
NGC 1560 is  very similar
to those derived {from} globular clusters.

The models based on the yields by WLW \& WW predict lower $r$
values than those based on the yields by Maeder (1992).
For a given model that fits $\mu_{IMF}$ and O, the yields by WLW \& WW
predict higher C/O, $\Delta Y/\Delta {\rm O}$, and $Z$/O values 
than the yields by
Maeder.
The closed-box models with continuous SFRs 
based on the yields by WLW \& WW can fit
the observational constraints provided by the well observed irregular
galaxies. In other words, the O-rich outflows that are required
by the yields of Maeder to fit the typical irregular galaxy are not
required by the yields of WLW \& WW.

For models with the same age the C/O, 
$\Delta Y/\Delta {\rm O}$, and $Z$/O ratios are almost independent 
of the $r$ value.

The fit between the C/O and $Z$/O 
ratios predicted by the closed-box  models with continuous SFRs
and the observational constraints is only fair.

It is possible to obtain lower $r$ values by adopting:
a) O-rich outflow models or
b) closed-box models  
with higher $M_{nb}$ values and  lower $M_{sub}$ values
(see Table 5).
Nevertheless, O-rich outflow models can be disregarded
for the typical irregular galaxy 
because they predict higher C/O and $Z$/O  values than
closed-box models and consequently larger differences with
the observed values.

A C/O determination and a better determination of $\mu_{IMF}$ is
needed for II Zw 33 to be able to discriminate among the different
models that we have computed for this object.

The dark matter mass fraction for the models computed in this paper
 amounts to about 40\%, part could be baryonic
(substellar) and part  non baryonic. This result implies that 
$M_{sub}$ is smaller than about 40\% and that the mass
fraction of non baryonic dark matter inside the Holmberg radius is also
smaller than about 40\%. 
For $r = 1.85$, the $r$ value derived {from} globular clusters, it follows that
for the typical irregular galaxy $M_{sub} = 26.6$ \% and $M_{nb} = 13.6$ \%
inside $R_H$.

By comparing  bursting SFR models
with continuous SFR models of the same age the differences in
the final abundance ratios
are very small. Consequently, it can be  said that the shape
of the SFR does not affect the results considerably. The largest differences
occur just after a burst: the C/O, $\Delta Y/\Delta {\rm  O}$, 
and  $Z$/O values decrease,
diminishing the differences of C/O and $Z$/O with the observed
values but increasing the differences of the $\Delta Y/\Delta {\rm  O}$
with observed value.

\acknowledgments 

Manuel Peimbert acknowledges several illuminating discussions with
Evan Skillman and Gerry Gilmore during the Symposium on Cosmic Chemical
Evolution held in NORDITA to honor Professor Bernard Pagel.
We also acknowledge a thorough reading of an earlier version of this
paper and several excellent suggestions by the referee Mario Mateo.
We made use of the NASA/IPAC Extragalactic Database (NED) which is operated
by the Jet Propulsion Laboratory, California Institute of Technology,
under contract with the National Aeronautics and Space Administration.
This work was partially supported by
DGAPA/UNAM through
project IN-100994.


\begin{thebibliography}{DUM}
\bibitem[]{}
Arimoto, N., Sofue, Y., \& Tsujimoto, T. 1996, PASJ, 48, 275
\bibitem[]{}
Bajaja, E., Huchtmeier, W.K., \& Klein, U. 1994, A\&A, 285, 385
\bibitem[]{}
Balkowski, C., Chamaraux, P., \& Weliachew, L. 1978, A\&A,
69, 263
\bibitem[]{}
Brinks, E., \& Klein, U. 1988, MNRAS, 231, 63
\bibitem[]{}
Broeils, A.H. 1992, A\&A, 256, 19
\bibitem[]{}
Burlak, A.N. 1996, Astron. Rep., 40, 621
\bibitem[]{}
Carigi, L. 1996, Rev. Mexicana Astron. Astrofis., 32, 179
\bibitem[]{}
Carigi, L., Col\I n, P., Peimbert, M., \& Sarmiento, A. 1995, ApJ, 445, 98 (CCPS)
\bibitem[]{}
Carigi, L., \& Peimbert, M. 1998, Rev. Mexicana Astron. Astrofis., submitted
\bibitem[]{}
D'Antona, F., \& Mazzitelli, I. 1995, ApJ, 456, 329
\bibitem[]{}
Esteban, C., \& Peimbert, M. 1995, A\&A, 300, 78
\bibitem[]{}
Esteban, C., Peimbert, M., Torres-Peimbert, S., \& Escalante, V. 1998, MNRAS, 
295, 401
\bibitem[]{}
Ferrano, F.R., Carretta, E., Bragaglia, A., Renzini, A., \& Ortolani, S.
1997, MNRAS, 286, 1012 (FCBRO)
\bibitem[]{}
Garnett, D.R., Skillman, E.D., Dufour, R.J., Peimbert, M., 
Torres-Peimbert, S., Terlevich, R., Terlevich, E., \& Shields, G.A. 1995,
ApJ, 443, 64
\bibitem[]{}
Garnett, D.R., Skillman, E.D., Dufour, R.J., \& Shields, G.A. 1997, 
ApJ, 481, 174
\bibitem[]{}
Gottesman, S.T., \& Weliachew, L. 1972, ApJ, 12, 63
\bibitem[]{}
---------. 1977, A\&A, 61, 523
\bibitem[]{}
Hindman, J.V. 1967, AuJPh, 20, 147
\bibitem[]{}
Israel, F.P. 1997, A\&A, 317, 65
\bibitem[]{}
Izotov, Y.I., Thuan, T.X., \& Lipovetski, V.A. 1997, ApJS, 108, 1
\bibitem[]{}
Kennicutt, R.C. 1998, in The Stellar Initial Mass Function, eds. G.
Gilmore \& D. Howell, (San Francisco:ASP Conference Series), 1
\bibitem[]{}
Kobulnicky, H.A., Skillman, E.D., Roy, J.R., Walsh, J.R., \& Rosa, M.R.
1997, ApJ, 477, 679
\bibitem[]{}
Kroupa, P., Tout, C.A., \& Gilmore, G. 1993, MNRAS, 262, 545 (KTG)
\bibitem[]{}
Kunkel, W.E., Demers, S., Irwin, M.J. 1996, AAS, 188.6504.
\bibitem[]{}
Kunkel, W.E., Demers, S., Irwin, M.J., \& Albert, L. 1997, ApJ, 488, L129
\bibitem[]{}
Lequeux, J. 1996, in the Interplay between Massive Star
Formation, the ISM and Galaxy Evolution, eds. D. Kunth,
B. Guiderdoni, M. Heydari-Malayeri, \& Trinh Xuan Thuan 
(Editions Frontieres), 105
\bibitem[]{}
Lequeux, J., Peimbert, M., Rayo, J., Serrano, A., 
\& Torres-Peimbert, S. 1979, A\&A, 80, 155
\bibitem[]{}
Lo, K.Y., Sargent, W.L.W., \& Young, K. 1993, AJ, 106, 507
\bibitem[]{}
Madden, S.C., Poglitsch, A., Geis, N., Stacey, G.J., \& Townes, C.H.
1997, ApJ, 483, 200
\bibitem[]{}
Maeder, A.  1992, A\&A, 264, 105
\bibitem[]{}
Maloney, P., \& Black, J.H. 1988, ApJ, 325, 389
\bibitem[]{}
Mateo, M. 1998, ARA\&A, 36, 435
\bibitem[]{}
Moore, B. 1994, Nature, 370, 629
\bibitem[]{}
Nomoto, K., Thielemann, F.K., \&  Yokoi, K. 1984, ApJ, 286, 644
\bibitem[]{}
Olive, K.A., Steigman, G., \& Skillman, E.D. 1997, ApJ, 483, 788
\bibitem[]{}
Pagel, B.E.J., \& Portinari, L. 1998, MNRAS, 298, 747
\bibitem[]{}
Pagel, B.E.J., \& Tautvai\v{s}ien\.e, G. 1998, MNRAS in press
\bibitem[]{}
Peimbert, M., Col\I n, P., \& Sarmiento, A. 1994, in
 Violent Star Formation, ed. G. Tenorio-Tagle,
(Cambridge University Press), 79
\bibitem[]{}
Peimbert, M., Torres-Peimbert, S., \& Ruiz, M.T. 1992, Rev. Mexicana Astron. Astrofis., 24, 155
\bibitem[]{}
Petrosian, A.R., Boulesteix, J., Comte, G., Kunth, D., \&
LeCoarer, E. 1997, A\&A, 318, 390
\bibitem[]{}
Piotto, G., Cool, A.M., \& King, I.R. 1997, AJ, 113, 1345 (PCK)
\bibitem[]{}
Renzini, A., \& Voli, M. 1981, A\&A, 94, 175
\bibitem[]{}
Richer, M.G., \&  McCall, M.L. 1995, ApJ, 445, 659
\bibitem[]{}
Salucci, P., \& Persic, M. 1997, in
Dark and Visible Matter in Galaxies, eds. M. Persic,
\& P. Salucci (San Francisco: ASP Conference Series), 1
\bibitem[]{}
Sandage, A., \& Tammann, G.A. 1975, ApJ, 196, 313
\bibitem[]{}
Schaller, G., Schaerer, D., Meynet, G., \& Maeder, A. 1992, A\&AS, 96, 269
\bibitem[]{}
Shostak, G.S. 1974, A\&A, 31, 97
\bibitem[]{}
Staveley-Smith, L., Davies, R.D., \& Kinman, T.D. 1992,
MNRAS, 258, 334 (SDK)
\bibitem[]{}
Tacconi, L.J., \& Young, J.S. 1987, ApJ, 322, 681
\bibitem[]{}
Tosi, M. 1996, ASP Conf. Ser. 98, {From} Stars to Galaxies:
The Impact of Stellar Physics on Galaxy Evolution,
eds. C. Leitherer, U.F. von-Alvensleben, \&  J. Huchra,
(San Francisco:ASP), 299
\bibitem[]{}
---------. 1998, preprint (astro-ph/9806266).
\bibitem[]{}
Walter, F., Brinks, E., Duric, N., \& Klein, U. 1997,
AJ, 113, 2031
\bibitem[]{}
Welch, D.L., McLaren, R.A., Madore, B.F., \& McAlary, C.W. 
1987, ApJ, 321, 162
\bibitem[]{}
Wilson, C.D. 1995, ApJ, 448, 97
\bibitem[]{}
Wilson, C.D., Welch, D.L., Reid, I.N., Saha, A., \& Hoessel, J. 1996
AJ, 111, 1106.
\bibitem[]{}
Woosley, S.E., Langer, N., \& Weaver, T.A. 1993, ApJ, 411, 823 (WLW)
\bibitem[]{}
Woosley, S.E., \& Weaver, T.A.  1995, ApJS 101, 181 (WW)
\end{thebibliography}
\end{document}